\documentclass[aps,prx,twocolumn,showpacs,superscriptaddress]{revtex4-2}
\usepackage{CJK,graphicx,setspace}
\usepackage{amsmath}
\usepackage{multirow}
\usepackage{tabularx}
\usepackage{booktabs}
\usepackage{color}
\usepackage{xcolor}
\usepackage{ulem}
\usepackage[colorlinks,linkcolor=blue,urlcolor=blue,anchorcolor=blue,citecolor=blue]{hyperref}
\begin{document}
\preprint{APS/123-QED}
\newcommand{\ccq}{\textcolor{orange}}
\newcommand{\qwy}{\textcolor{red}}

\begin{CJK*}{GBK}{song}
\title{\mbox{Electronic structures and superconductivity in Nd-doped La$_3$Ni$_2$O$_7$ }}

 
\author{Cui-Qun Chen}
\thanks{These two authors contributed equally to this work.}
 
\author{Wenyuan Qiu}
\thanks{These two authors contributed equally to this work.}

\author{Zhihui Luo}

\author{Meng Wang}

\author{Dao-Xin Yao}
\email{yaodaox@mail.sysu.edu.cn}

\affiliation{Center for Neutron Science and Technology, Guangdong Provincial Key Laboratory of Magnetoelectric Physics and Devices, State Key Laboratory of Optoelectronic Materials and Technologies, School of Physics, Sun Yat-Sen University, Guangzhou 510275, China}

\begin{abstract}

The recent discovery of high-$T_c$ superconductivity in Ruddlesden-Popper (RP) nickelates has motivated extensive efforts to explore higher $T_c$ superconductors. Here, we systematically investigate Nd-doped La$_3$Ni$_2$O$_7$ using density functional theory (DFT) and renormalized mean-field theory (RMFT). DFT calculations reveal that both the lattice constants and interlayer spacing decrease upon Nd substitution, similar to the effect of physical pressure. However, the in-plane Ni-O-Ni bond angle evolves non-monotonically with doping, increasing to a maximum at 70\% ($\sim$ 2/3) Nd doping level and then falling sharply at 80\%, which leads to a reduction in orbital overlap. Moreover, Nd doping has a more pronounced effect on the Ni-$d{_{z^2}}$ orbital, demonstrating an orbital-dependent effect of rare-earth substitution. Through the bilayer two-orbital $t-J$ model, RMFT analysis further shows an $s\pm$-wave pairing symmetry, with $T_c$ rising to a maximum at about 70\% Nd substitution before declining, in agreement with the transport measurements. The variation in $T_c$ can be traced to the competition between continuously enhanced interlayer superexchange coupling $J_\perp^z$ and a gradual decrease in particle density. These results highlight the delicate interplay among structural tuning, orbital hybridization, and superconductivity, providing important clues to design higher-$T_c$ RP nickelate superconductors.

\end{abstract}

\maketitle
\end{CJK*}
\noindent{\bf 1 Introduction}
\vspace{0.15in}

\noindent Since the discovery of superconductivity above liquid-nitrogen temperature in the Ruddlesden-Popper (RP) phase bilayer nickelate La$_3$Ni$_2$O$_7$~\cite{bilayernature}, nickelate superconductors have drawn broad attention in condensed matter physics, sparking intensive experimental and theoretical investigations~\cite{bilayermodel,huomodulation,zhang2023electronic,lechermann2023electronic,luo2024high,wu2024superexchange,  cpl_41_7_077402,shilenko2023correlated,yang2023interlayer, zhang2023trends,PhysRevLett.131.206501,shen2023effective,oh2023type,YangF2023,liao2023electron,WangQH2023,PhysRevB.110.235155,kaneko2024pair,ouyang2024absence,heier2024competing,zhang2024structural,zhang2024electronic,tian2024correlation,ryee2024quenched,zhang2024strong, ni_spin_2025,lu2024interlayer,qu2024bilayer,yang2024strong, fan2024superconductivity,   sakakibara2024possible, cao2024flat, jiang2024pressure,chen2025charge,yang_orbital-dependent_2024,Hou_2023, PhysRevB.110.134520, wang_pressure-induced_2024, liu_electronic_2024,PhysRevLett.134.076001, shao2024possiblehightemperaturesuperconductivitydriven}. Subsequent studies have reported superconductivity in a variety of RP phase nickelates, including trilayer La$_4$Ni$_3$O$_{10}$ with zero-resistivity signatures, hybrid RP-phase nickelate La$_5$Ni$_3$O$_{11}$, as well as thin-film La$_3$Ni$_2$O$_7$ and La$_5$Ni$_3$O$_{11}$ that exhibit superconductivity even at ambient pressure~\cite{zhu_superconductivity_2024,zhang2025superconductivity,ko_signatures_2025,zhou_ambient-pressure_2025,experimental,PhysRevB.110.L180501,PhysRevB.110.014503,zhang2025pairingmechanismsuperconductivitypressurized}. A key challenge in this field is how to further enhance the superconducting transition temperature ($T_c$). Chemical pressure induced by rare-earth substitution offers an effective method to modulate $T_c$, as exemplified by the lanthanide-radius-dependent $T_c$ trend observed in iron-based superconductors~\cite{RevModPhys.83.1589}. Theoretically, it has been proposed that replacing La in La$_3$Ni$_2$O$_7$ with other rare-earth elements -- for instance, Sm$_3$Ni$_2$O$_7$ -- could potentially double the $T_c$~\cite{cpl_41_8_087401}. However, realizing full rare-earth substitution remains experimentally challenging.

Recently, the successful synthesis of La$_2$SmNi$_2$O$_7$ has advanced the superconducting transition of nickelates into the 90 K regime~\cite{li2025ambientpressuregrowthbilayer}. Soon after, Nd-doped La$_3$Ni$_2$O$_7$ further pushed $T_c$ to 93 K from the electronic transport measurement, in which radio-frequency measurements even found the signature of superconductivity at 98 K ~\cite{private}. Interestingly, experimental observations indicate that the evolution of $T_c$ with doping concentration may not be monotonic, in contrast to the intuition derived from theoretical studies on full La substitution~\cite{cpl_41_8_087401}. Therefore, understanding how doping concentration modulates the superconducting $T_c$ in La$_3$Ni$_2$O$_7$ systems has emerged as a significant question.

The RP-phase La$_3$Ni$_2$O$_7$ consists of alternating stacks of two NiO$_2$ planes, where each Ni atom forms an octahedron with surrounding oxygen atoms, giving rise to a quasi-two-dimensional structure. The Ni ions in La$_3$Ni$_2$O$_7$ exhibit a nominal valence state of 3$d^{7.5}$ with the Ni-$d_{z^2}$ and $d_{x^2-y^2}$ orbitals near the Fermi level being approximately half-filled and quarter-filled, respectively. At ambient pressure, La$_3$Ni$_2$O$_7$ crystallizes in an orthorhombic space group $Amam$, characterized by unequal lattice constants $a \neq b$ and tilted octahedra that cause the out-of-plane Ni-O-Ni bond angle to deviate from 180$^\circ$~\cite{li_identification_2025}. Upon external high pressure, the system undergoes a structural phase transition in which the in-plane lattice constants $a$ and $b$ converge, and the out-of-plane Ni-O-Ni bond angle approaches 180$^\circ$. The critical pressure for this structural transition depends sensitively on the choice of the rare-earth ion, which also determines the optimal pressure for maximal $T_c$, 
underscoring the role of chemical pressure in governing the electronic structures and superconducting properties of nickelates~\cite{samanta2025inplanenionibondangles}.

\begin{figure*}
\noindent \begin{centering}
\includegraphics[width=2\columnwidth,height=2\columnwidth,keepaspectratio]{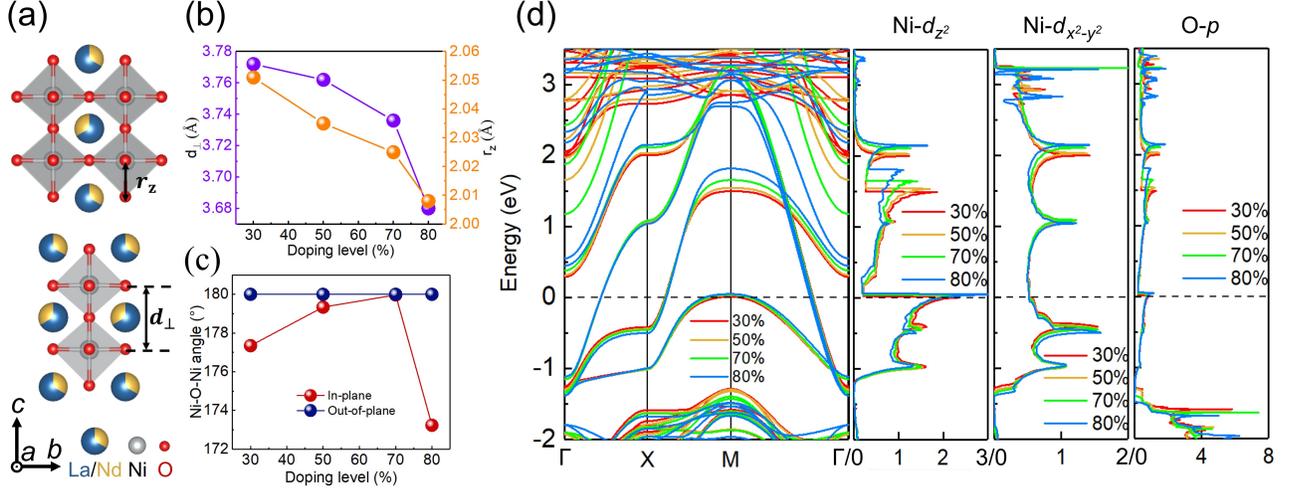}\caption{ (a) Crystal structure of Nd-doped La$_{3}$Ni$_2$O$_7$. The blue, yellow, red
 and grey balls denote La, Nd O, and Ni atoms. $d_\perp$ and $r_z$ indicate the interlayer Ni distance and the apical Ni-O bond length respectively. (b) Interlayer distance and apical Ni-O bond length versus doping level. (c) Ni-O-Ni angle versus doping level. (d) Band structures and projected DOS of Nd-doped La$_3$Ni$_2$O$_7$ at varying doping level under high pressure. 
  \label{fig1}}
\par\end{centering}
\end{figure*}

In this work, we systematically investigate the structural, electronic, and superconducting properties of Nd-doped La$_3$Ni$_2$O$_7$ using first-principles density functional theory (DFT) in combination with the renormalized mean-field theory (RMFT). Within DFT calculations, we analyze the evolution of bond angles and bond lengths under different Nd doping concentrations, and construct a bilayer two-orbital model and a high energy 11-orbital model based on the electronic structures. Our results reveal that Nd doping exerts an orbital-dependent influence on the electronic structure, with the hopping integrals exhibiting a non-monotonic dependence on doping concentration. RMFT calculations further indicate that the system favors an $s\pm$-wave pairing symmetry similar to that of undoped La$_3$Ni$_2$O$_7$, and that $T_c$ also evolves non-monotonically with doping, reaching its maximum at about 70\% ($\sim$2/3) Nd substitution. Our results provide a microscopic understanding of how rare-earth substitution and site occupancy influence superconductivity in RP phase nickelates, thereby offering valuable guidance for designing higher-$T_c$ nickelate superconductors.

\vspace{0.2in}
\noindent{\bf 2 Model and methods}
\vspace{0.15in}


DFT calculations were performed using Vienna ab initio simulation package (VASP)~\citep{VASP1,VASP2}, in which the projector augmented wave (PAW) \cite{PAW1,PAW2} method with a 500 eV plane-wave cutoff is applied. 
The generalized gradient approximation (GGA) of Perdew-Burke-Ernzerhof form (PBE) exchange correlation potential is adopted \cite{PhysRevLett.77.3865}.
The total energy convergence criterion was set to $10^{-7}$ eV and a $\Gamma$-centered $19\times19\times19$ Monkhorst Pack k-mesh grid is used for primitive cell of $I4/mmm$ phase.
In structural relaxations, we adopted the experimental refined lattice constants of Nd-doped $\mathrm{La_3Ni_2O_7}$ at 37 GPa~\cite{private} with the convergence criterion of force set to 0.001\ eV/{\rm \AA}. Correlation effect is considered with an effective $U_{eff}=3.5$ eV~\cite{yang_orbital-dependent_2024}.
Guided by the previous scanning transmission electron microscopy study which indicates a preferential incorporation of dopant atoms at the outer-layer R sites, our calculations systematically substitute Nd at the outer-layer La sites until full occupancy is reached, with the remaining dopants then occupying the inner-layer sites~\cite{li2025ambientpressuregrowthbilayer}.
To obtain the projected tight-binding models, we further performed Wannier downfolding  as implemented in WANNIER90 package~\cite{w90}, in which good convergence was reached.


For RMFT calculation, we construct the bilayer two-orbital $t-J$ model as follows:
\begin{align}
    H =& H_t +H_J, \nonumber \\
    H_t =& \sum_{i\alpha\sigma}\varepsilon_{\alpha}\hat{c}^{\dagger}_{i\alpha\sigma}\hat{c}_{i\alpha\sigma}+\sum_{ij,\alpha\beta,\sigma}t_{ij}^{\alpha\beta}(\hat{c}^{\dagger}_{i\alpha\sigma}\hat{c}_{j\beta\sigma}+h.c.),
        \nonumber \\
    H_{J} =& 
        J_{\bot}^{z}\sum_{i}S_{iz_1}.S_{iz_2} + J_{||}^{x}\sum_{<ij>}^{\alpha=x_1,x_2}S_{i\alpha}.S_{j\alpha} \nonumber\\
        &-J_H\sum_{i}^{\alpha\beta=x_1z_1,x_2z_2}S_{i\alpha}.S_{i\beta}. 
        \label{hamiltonian}
\end{align} 
Here, $H_t$ is the tight-binding model. $i/j$, $\sigma$ and $\alpha/\beta$ denote the indices of site (including the two layers), spin, and orbital ($d_{x^2-y^2}$ or $d_{z^2}$), respectively. $\varepsilon_{\alpha}$ represents the on-site energy of orbital $\alpha$. $J_{\bot}^z$, $J_{||}^x$ and $J_H$ in $H_J$ represent interlayer $d_{z^2}$, in-plane $d_{x^2-y^2}$ superexchange couplings and Hund's interaction, respectively. 
The $J_{\bot}^z$ can be estimated through Exact Diagonalization (ED) with an effective Hamiltonian while we set $J_{||}^x=J_{\bot}^z/2$ and $J_H=1$ eV during our calculation.
The symbol $z_1/x_1$ ($z_2/x_2$) represents orbital $d_{z^2}/d_{x^2-y^2}$ of two layers.

Within the RMFT framework, the $t-J$ model in Eq.\ref{hamiltonian} is treated by applying a conventional mean-field decoupling to the spin-exchange interaction, leading to the order parameters $\chi_{ij}^{\mu\nu}=\langle c^{\dagger}_{i\mu\uparrow} c_{j\nu\uparrow}\rangle$ and $\Delta^{\mu\nu}_{ij} = \langle c^{\dagger}_{i\mu\uparrow} c^{\dagger}_{j\nu\downarrow} \rangle$.
Furthermore, the Gutzwiller renormalization factors $g_t$ and $g_J$ are introduced into $H_t$ and $H_J$, which reflect the strong renormalization from correlations~\cite{gutz}.
Finally, the obtained quadratic Hamiltonian is solved self-consistently while keeping the electron densities fixed at a certain level.
The eventual pairing symmetry can be determined from the phase structure of various pairing bonds $\Delta^{\mu\nu}_{ij}$, and $T_c$ is defined as the temperature where the energy gap from the interlayer $d_{z^2}$ orbital falls to two-thirds of its value at 0 K.
A detailed description of the RMFT approach is given in Ref.~\cite{luo2024high}.

\begin{table}[t]
\caption{Site energies $\epsilon_{x/z}$ and tight-binding parameters $t_{[i,j,k]}^{x/z}$ of bilayer two-orbital model for Nd-doped $\mathrm{La_3Ni_2O_7}$. $t_{[i,j,k]}^{x/z}$ denotes the hopping term that is connected by $[0,0,0]$-$[i,j,k]$ bond within $d_{x^2-y^2}/d_{z^2}$ orbital, while  $t^{xz}_{[i,j,k]}$ denotes the hybridization between them. Note that the vertical interlayer distance is assigned as 1 and all parameters are in unit of eV. The symmetrically equivalent terms are not shown for clarity. }
\begin{onehalfspace}\label{tab:tb2orb}
\noindent\begin{centering}
\begin{tabular}{ccccccc}
\hline \hline 
Doping level & $\epsilon_{d_{x^2-y^2}}$ & $\epsilon_{d_{z^2}}$ & $t_{[100]}^{x}$ & $t_{[100]}^{z}$ & $t_{[110]}^{x}$ & $t_{[110]}^{z}$  \tabularnewline
\hline 
30\%  & 1.000 & 0.357 & -0.541 & -0.121 & 0.066 & -0.021 \tabularnewline

50\% & 0.984 & 0.394 & -0.542 & -0.114 & 0.065 & -0.012 \tabularnewline

70\% & 0.962 & 0.422 & -0.546 & -0.127 & 0.067 & -0.018 \tabularnewline

80\% & 0.902 & 0.447 & -0.544 & -0.149 & 0.068 & -0.021 \tabularnewline
\hline \hline 
Doping level & $t_{[200]}^{x}$ & $t_{[200]}^{z}$ & $t_{[100]}^{xz}$ & $t_{[200]}^{xz}$ & $t_{[001]}^{z}$ &  $t_{[101]}^{xz}$ \tabularnewline
\hline 
30\% & -0.074 & -0.002 & 0.264 & 0.040 & -0.726 & -0.024 \tabularnewline
 
50\% & -0.076 & -0.003 & 0.268 & 0.040 & -0.755 & -0.028 \tabularnewline

70\% & -0.078 & -0.002 & 0.276 & 0.039 & -0.808 & -0.035 \tabularnewline
 
80\% & -0.079 & -0.011 & 0.280 & 0.039 & -0.847 & -0.041 \tabularnewline
\hline \hline 
\end{tabular}
\par\end{centering}
\end{onehalfspace}
\end{table}

\vspace{0.2in}
\noindent{\bf 3 Results}
\vspace{0.15in}

\noindent{\bf {\small 3.1 Crystal and electronic structures}}
\vspace{0.1in}

Crystal structure of Nd-doped La$_3$Ni$_2$O$_7$ under high pressure is displayed in Fig.\ref{fig1}(a). Synchrotron X-ray diffraction measurement indicates a structural phase transition from orthorhombic to tetragonal structure with increasing pressure, similar to the undoped La$_3$Ni$_2$O$_7$ \cite{li_identification_2025}. Here, based on the experimental refined lattice constants, we perform DFT calculations on the Nd-doped La$_3$Ni$_2$O$_7$ of $I4/mmm$ phase. Owning to the smaller ionic radius of Nd, the increase in the doping concentration leads to a reduction in the lattice constants, the apical Ni-O bond length, and the interlayer distance, as displayed in Fig.\ref{fig1}(b). However, Figure \ref{fig1}(c) shows that the in-plane Ni-O-Ni angle gradually increases as the doping level increases from 30\% to 70\%, followed by a sudden drop at 80\% doping concentration. The increase in in-plane Ni-O-Ni bond angle from 30\% to 70\% doping concentration can be attributed to the progressive approach toward full occupancy of the outer-layer R site by Nd atoms. Conversely, the reduction of the in-plane bond angle at 80\% doping level results from the incorporation of Nd into the inner layer, leading to a variation in the atomic radii of the inner-layer sites.
Previous studies on undoped La$_3$Ni$_2$O$_7$ indicated that the in-plane Ni-O-Ni bond angle decreases with increasing pressure, a trend that is consistent with the $T_c$ under high pressure~\cite{verraes2025evidencestronglycorrelatedsuperconductivity,samanta2025inplanenionibondangles}. Therefore, the reduction of in-plane Ni-O-Ni bond angle at 80\% doping concentration may potentially affect $T_c$.

In Fig.\ref{fig1}(d), we present the electronic structures at various doping levels under high pressure. The electronic structures near the Fermi level are primarily dominated by the Ni-$e_g$ orbitals, which is similar to the undoped La$_3$Ni$_2$O$_7$~\cite{bilayermodel}. As the doping concentration increases, the antibonding state of the $d_{z^2}$ orbital shifts towards higher energy, leading to a corresponding enlargement of the energy separation between the bonding and antibonding states of the Ni-$d_{z^2}$ orbital, implying the enhancement of the interlayer $d_{z^2}$ orbital couplings. The bonding state of $d_{z^2}$ orbital, however, exhibits merely a minor upward shift with rising doping levels, leading to a slight enlargement of the hole pocket at the Fermi surface (FS). 
From the band structures, we see that although Nd doping similarly leads to a reduction in lattice constants, in contrast to the physical pressure effect, the bandwidth of the $d_{x^2-y^2}$ orbital does not exhibit significant broadening. Instead, the effect of Nd doping is predominantly manifested in the $d_{z^2}$ orbital electronic structure, exhibiting an orbital-dependent effect.

\begin{figure}[t]
\noindent \begin{centering}
\includegraphics[width=1.0\columnwidth,height=1.0\columnwidth,keepaspectratio]{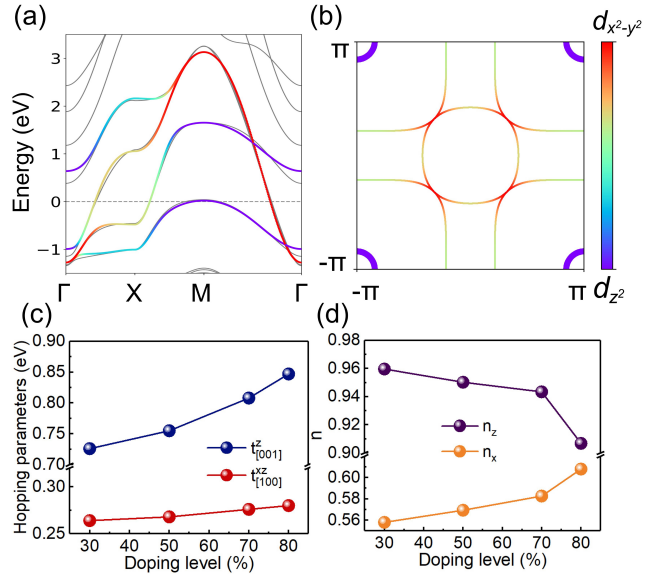}\caption{(a) Band structure and (b) FS of bilayer two-orbital model for 70\% Nd-doped La$_3$Ni$_2$O$_7$. The color bar indicates the orbital weight of Ni-$d_{z^{2}}$ and $d_{x^{2}-y^{2}}$. The grey lines in (a) are band structures from DFT. (c)-(d) Variation with doping level of (c) $t_{[001]}^z$ and $t_{[100]}^{xz}$, (d) density of $d_{z^2}$ and $d_{x^2-y^2}$ orbitals.\label{fig2}}
\par\end{centering}
\end{figure}

\vspace{0.2in}
\noindent{\bf {\small 3.2 Electronic model}}
\vspace{0.1in}

To gain more insights into electronic properties that are directly relevant to superconductivity,  we further perform Wannier downfolding on the Ni-$e_g$ orbital based on the DFT electronic structures, which allows us to construct a bilayer two-orbital model written as shown in $H_{t}$ in Eq.~\ref{hamiltonian}.


\begin{table}[t]
\caption{Site energies $\epsilon^{d/p}_{x/z}$ and tight-binding parameters of eleven-orbital model for Nd-doped La$_3$Ni$_2$O$_7$ in unit of eV. $d_{x/z}$ and $p_{x/y/z}$ indicate the  $d_{x^2-y^2}/d_{z^2}$ and $p_x/p_y/p_z$ orbitals. The symmetrically equivalent terms are not shown for clarity. }
\begin{onehalfspace}\label{tab:tb11orb}
\noindent\begin{centering}
\begin{tabular}{ccccccc}
\hline \hline 
Doping level & $d_x$-$p_{x}$ & $d_z$-$p_{x}$ & $d_z$-$p_z$ & $d_z$-$p_{z\prime}$ & $p_x$-$p_y$ & $p_z$-$p_{x/y}$  \tabularnewline
\hline 
30\% & 1.761 & -0.822 & 1.877 & 1.432 & 0.576 & 0.478 \tabularnewline

50\% & 1.817 & -0.827 & 1.896 & 1.452 & 0.605 & 0.512 \tabularnewline

70\% & 1.849 & -0.832 & 1.983 & 1.521 & 0.649 & 0.544 \tabularnewline

80\% & 1.795 & -0.879 & 1.975 & 1.495 & 0.602 & 0.512 \tabularnewline

\hline \hline 
Doping level & $p_{z^\prime}$-$p_{x}$ & $\epsilon^d_x$ & $\epsilon^d_z$ & $\epsilon^p_{x}$ & $\epsilon^p_z$  & $\epsilon^p_{z^\prime}$ \tabularnewline
\hline 

30\% & 0.494 & -1.259 & -1.626 & -5.304 & -4.771 & -3.807 \tabularnewline

50\% & 0.499 & -1.303 & -1.598 & -5.367 & -4.828 & -3.967 \tabularnewline

70\% & 0.510 & -1.326 & -1.572 & -5.503 & -4.981 & -4.380 \tabularnewline

80\% & 0.488 & -1.342 & -1.553 & -5.470 & -4.628 & -4.426 \tabularnewline
\hline \hline 
\end{tabular}
\par\end{centering}
\end{onehalfspace}
\end{table}

We consider hoppings up to the third-nearest neighbor in order to accurately describe the low-lying state of our DFT results, which yields 12 parameters for each doping level, as listed in Table \ref{tab:tb2orb}. The band structures and FSs obtained from the bilayer two-orbital model are presented in Fig.\ref{fig2}(a-b). Results for the optimal 70\% doping with the highest $T_c$ are shown in the main text, whereas others are included in the Supplementary Material.
Our two-orbital bilayer model reveals that different doping concentrations consistently yield three Fermi pockets on the FS, two electron pockets ($\alpha$ and $\beta$) and one hole pocket ($\gamma$), which is analogous to the undoped La$_3$Ni$_2$O$_7$. The distinction, however, lies in the significant enhancement of both the nearest-neighbor hopping of the $d_{x^2-y^2}$ orbital and the interlayer hopping of the $d_{z^2}$ orbital, which can be attributed to the reduction in lattice constants as well as interlayer distance ($d_\perp$) induced by doping. Specifically, as the doping concentration increases, the site energy of the $d_{x^2-y^2}$ orbital ($\epsilon_{d_{x^2-y^2}}$) decreases from 1.000 to 0.902, while the interlayer hopping of the $d_{z^2}$ orbital $t^z_{[001]}$ increases from -0.726 to -0.847, implying a gradual strengthening of interlayer coupling with higher doping levels.

The bilayer two-orbital model reveals a redistribution of electrons between the $d_{x^2-y^2}$ and $d_{z^2}$ orbitals with Nd doping. As doping level rises from 30\% to 80\%, the density of $d_{x^2-y^2}$ orbital increases from 0.558 to 0.608, while the $d_{z^2}$ orbital density decreases from 0.960 to 0.906 [as shown in Fig.\ref{fig2}(d)], demonstrating the significant electronic reorganization induced by Nd doping in La$_3$Ni$_2$O$_7$.

\begin{figure}
\noindent \begin{centering}
\includegraphics[width=1.0\columnwidth,height=1.0\columnwidth,keepaspectratio]{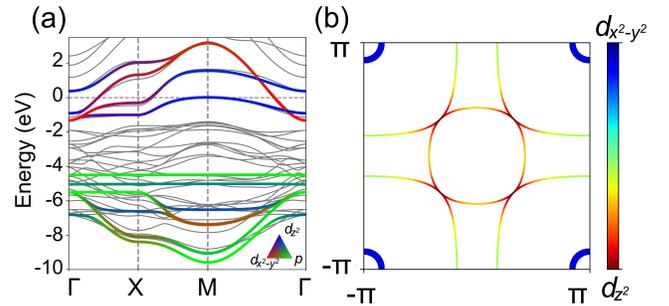}\caption{(a) Band structure and (b) FS of eleven-orbital model for 70\% Nd-doped La$_3$Ni$_2$O$_7$. The blue, red, and green colors in (a) denote orbital weight of Ni-$d_{z^{2}}$, Ni-$d_{x^{2}-y^{2}}$ and O-$p$ orbitals. The color bar in (b) indicates the orbital weight of Ni-$d_{3z^{2}-r^{2}}$ and $d_{x^{2}-y^{2}}$. The grey lines in (a) are band structures from DFT. \label{fig3}}
\par\end{centering}
\end{figure}

To account for the physics of O-$p$ orbitals as ligands mediating between Ni-$d$ orbitals, we further perform Wannier downfolding on Ni-$e_g$ and O-$p$ orbitals, which is formulated as a higher-energy 11-orbital model.
The basis is $\Psi=(d_{x^2-y^2},d_{z^2},d_{x^2-y^2},d_{z^2},p_x,p_y,p_x,p_y,p_z,p_z^\prime,p_z^\prime)$, as in the 11-orbital model for undoped La$_3$Ni$_2$O$_7$\cite{bilayermodel}. In this model, it is sufficient to consider 7 hopping parameters to fit the DFT band structures, including nearest-neighbor $d-p$ and $p-p$ hoppings. The specific parameters for each doping level are listed in Table.\ref{tab:tb11orb}. The resulting electronic band structures and FS are shown in Fig.\ref{fig3}. For clarity and simplicity, only the results for 70\% doping with highest $T_c$ are presented in the main text, while those for other doping concentrations are provided in the Supplementary Material. The 11-orbital model covers an energy range from -9 to 3.5 eV, with the contributions near the Fermi level dominated primarily by Ni-$e_g$ orbitals (red and blue), and the lower-energy regions characterized mainly by O-$p$ orbitals (green), similar to undoped La$_3$Ni$_2$O$_7$.

From the hopping parameters listed in Table.\ref{tab:tb11orb}, it can be observed that although the parameters generally increase with rising Nd-doping concentration, most of them exhibit a sudden decrease at the 80\% doping level. This phenomenon may be attributed to the abrupt incorporation of a substantial amount of Nd atoms into the inner layers at 80\% doping. On the other hand, the in-plane Ni-O-Ni bond angle deviates further from 180$^\circ$ at the 80\% doping level, thereby reducing the overlap between the in-plane $d_{x^2-y^2}$ and $p_x/p_y$ orbitals and consequently suppressing hopping integrals of $d_{x^2-y^2}$ and $p_x/p_y$ orbitals. This comparison indicates that the influence of Nd doping on the electronic properties of La$_3$Ni$_2$O$_7$ is not monotonic. In particular, when Nd atoms occupy the inner-layer sites in large quantities, it may not exert a further beneficial effect on the system's superconductivity.

\vspace{0.2in}
\noindent{\bf {\small 3.3 RMFT results}}
\vspace{0.1in}

\begin{figure}[t]
\noindent \begin{centering}
\includegraphics[width=1.0\columnwidth,height=1.0\columnwidth,keepaspectratio]{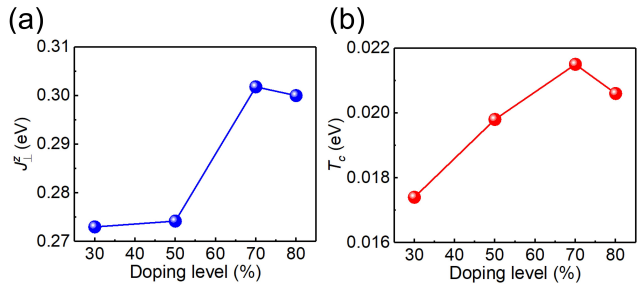}\caption{Variation with doping level of (a) interlayer $d_{z^2}$ superexchange coupling and (b) transition temperature $T_c$. \label{fig4}}
\par\end{centering}
\end{figure}

To advance the RMFT calculations, we next estimate the interlayer superexchange coupling of $d_{z^2}$ orbital ($J_{\bot}^z$).
A typical approximation for estimating superexchange coupling is $J\sim$ $4t^2/U$. 
In this context, $J_\bot^z$ will increase rapidly and $T_c$ would rise from doping level of 30\% to 80\% accordingly.
However, the experimental transport measurements show that $T_c$ reaches a peak at the doping level of 70\% and then decreases slightly ~\cite{private}. 
To understand this phenomenon, it is necessary to consider the effect of the decrease in $n_z$ when estimating $J_\bot^z$.
By constructing a 5-orbital O-$p_z$-Ni-$d_{z^2}$ simplified model based on the 11-orbital model above, ED can give a value of $J_\bot^z$ by the energy difference between the ground state and the first excited state with $U=8$ eV.
More details about ED can be found in Ref.~\cite{wu2024superexchange}.
As $n_z$ decreases, the double counting term $E_{dc}=\frac{1}{3}(U+2U^{\prime}-J_H)(n_d^0-0.5)$~\cite{Held01112007} in the 5-orbital model with $U=8$ eV, $U^{\prime}=U-2J_H$ and $J_H=0.1U$, results in the reduction of $J_{\bot}^z$, while the increase in hopping between Ni-$d_{z^2}$ and O-$p_z$ will raise $J_{\bot}^z$.
Thus $J_{\bot}^z$ reaches its largest value at doping level of 70\% 
and $T_c$ increases from doping level of 30\% to 70\%, followed by a slight decrease at 80\%, as shown in Figs.~\ref{fig4} (a) and (b), respectively. 
The evolution of $T_c$ closely follows that of $J_\perp^z$ and the overall trend is consistent with the experimental results~\cite{private}.
%

\begin{figure}[t]
\noindent \begin{centering}
\includegraphics[width=1.0\columnwidth,height=1.0\columnwidth,keepaspectratio]{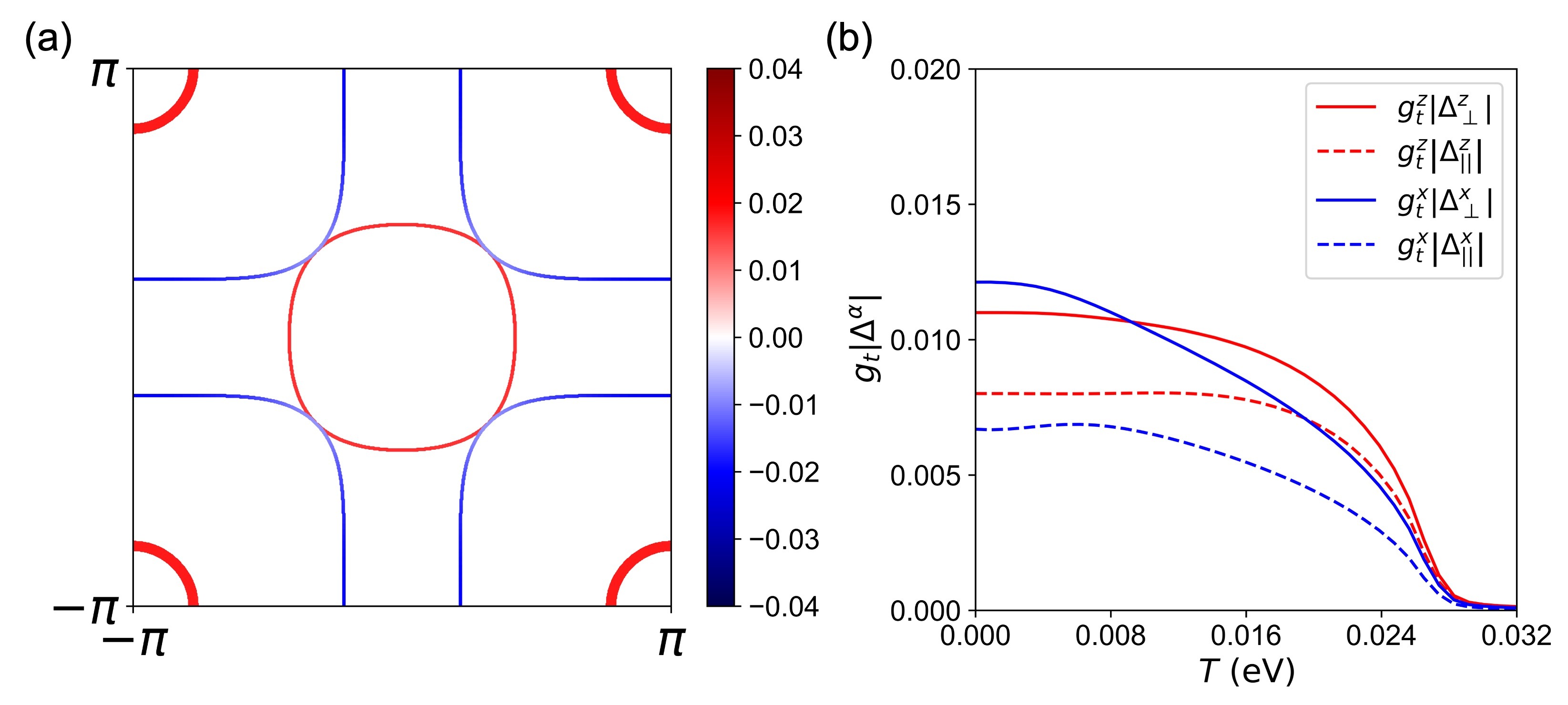}\caption{(a) The projection of energy gap on FS and (b) evolution of the energy gap with temperature for different pairing bonds at 70\% Nd-doping level. Here $g_t^{z/x}$ denotes the renormalization factor for $d_{z^2}/d_{x^2-y^2}$ orbital, and $\Delta_{\bot/||}^{z/x}$ is the interlayer/in-plane pairing bond for $d_{z^2}/d_{x^2-y^2}$ orbital.\label{fig5}}
\par\end{centering}
\end{figure}

Figure~\ref{fig5} (a) represents the projection of energy gap at $T=0$ K on FS for the optimal 70\% Nd doping level.
The pairing symmetry is clearly $s_\pm$-wave, characterized by the opposite signs of energy gap on the $\alpha$ / $\gamma$ and the $\beta$ pockets.
The $s_\pm$-wave pairing symmetry persists over the entire studied doping range.
Figure~\ref{fig5}(b) presents the temperature dependence of the energy gap associated with different pairing bonds at 70\% doping level.
For $d_{z^2}$ orbital pairing, the contribution from interlayer pairing is obviously larger than the one from in-plane pairing.
For the $d_{x^2-y^2}$ orbital pairing, the in-plane pairing energy gap is the smallest among all branches shown in Fig.~\ref{fig5}(b), whereas the interlayer pairing is initially larger but becomes smaller than that of the $d_{z^2}$ orbital pairing.
Since the $d_{z^2}$ orbital density is close to half-filling, the renormalization effect on energy gap becomes pronounced, as the Gutzwiller factors depend on orbital density.
As a result of the strong renormalization effect, the contributions to the energy gap from $d_{z^2}$ and $d_{x^2-y^2}$ pairing are very close.
The results for other Nd doping levels are provided in the Supplementary Material (Fig. S3), which shows that at 80\% doping, the energy gap associated with the $d_{z^2}$ orbital becomes dominant owing to the sharp decrease in $n_z$.

\vspace{0.2in}
\noindent{\bf 4 Discussion and conclusion}
\vspace{0.15in}

Our comprehensive investigation into Nd-doped La$_3$Ni$_2$O$_7$ reveals several key insights into the doping-dependent electronic structure and its implications for superconductivity in this bilayer nickelate system.

The orbital-dependent response to Nd doping represents a crucial distinction from the pressure effects observed in undoped La$_3$Ni$_2$O$_7$. While both chemical pressure via Nd doping and physical pressure reduce the lattice constants, their impacts on the electronic structures diverge significantly. Unlike the uniform bandwidth enhancement typically induced by physical pressure, Nd doping predominantly affects the $d_{z^2}$ orbital. 
The non-monotonic evolution of hopping integrals with doping concentration, particularly the anomalous suppression at 80\% Nd doping, reveals a complex interplay between chemical pressure and local structural distortions. The sudden decrease in hopping parameters at high doping levels coincides with both the incorporation of Nd into inner-layer sites and the deviation of Ni-O-Ni bond angles from 180$^\circ$. This correlation suggests that the doping effects are not merely governed by simple chemical pressure, but involve more subtle changes in the local bonding environment that affect orbital overlaps, particularly between in-plane Ni-$d_{x^2-y^2}$ and O-$p_x/p_y$ orbitals.

The transport measurements reveal an uncommon $T_c$ variation trend with increasing Nd doping level, in which $T_c$ slightly decreases after Nd doping of 70\%, although the radio-frequency measurements may give different evaluations of $T_c$~\cite{private}.
Within our RMFT framework, the $T_c$ obtained from the $t-J$ model in Eq.\ref{hamiltonian} is highly sensitive to the interlayer $d_{z^2}$ superexchange coupling $J_{\bot}^z$.
Another factor influencing $T_c$ in RMFT is orbital density.
The existence of Gutzwiller factors $g_t$ and $g_J$ will strongly depress half-filling states. 
But for the states away from half-filling, the decrease of orbital density will lower $T_c$. 
From Nd doping level of 30\% to 70\%, since the variation of orbital density is tiny, the enhancement of $T_c$ in this range is attributed primarily to the increase of $J_{\bot}^z$.
When Nd doping level reaches 80\%, $d_{z^2}$ orbital density sharply decreases, which could cancel the effect of the increase in hopping between Ni-$d_{z^2}$ and O-$p_z$ orbitals, and results in a slight reduction in $J_\bot^z$ caused by the $E_{dc}$.
Thus, $T_c$ can increase initially and then show a slight decrease at 80\%.

In summary, we systematically study the effect of different Nd doping levels in La$_3$Ni$_2$O$_7$ on crystal structure, electronic properties, pairing symmetry and $T_c$ based on DFT combined with RMFT. 
Our DFT calculations indicate that the chemical pressure induces a uniform reduction in bond lengths, while the in-plane Ni-O-Ni bond angle exhibits a non-monotonic trend. The electronic structures and bilayer two-orbital model indicate the orbital-dependent effect of Nd doping.
Then we use ED to estimate the $J_\bot^z$, with $J_{||}^x=J_\bot^z/2$ and investigate the pairing symmetry and $T_c$ with RMFT.
Our results show that $s\pm$ pairing symmetry does not change and $T_c$ could increase from doping level of 30\% to 70\% before a slight decrease at 80\%, providing guidelines for designing higher-$T_c$ nickelate superconductors.

The evolution of electronic properties and superconductivity with Nd doping in La$_3$Ni$_2$O$_7$ underscores the complex interplay among chemical pressure, local structure, and orbital-dependent correlations. These results not only advance our understanding of the superconducting mechanism in nickelates but also demonstrate chemical doping as a powerful tool for selectively tuning specific orbital degrees of freedom in correlated electron systems.

\begin{acknowledgments}
This project was supported by NSFC-12494591, NSFC-92165204, NSFC-92565303, NKRDPC-2022YFA1402802, CAS Superconducting Research Project (SCZX-0101), Guangdong Provincial Key Laboratory of Magnetoelectric Physics and Devices (2022B1212010008), Guangdong Fundamental Research Center for Magnetoelectric Physics (2024B0303390001), and Guangdong Provincial Quantum Science Strategic Initiative (GDZX2401010).

\end{acknowledgments}


\bibliography{lno327}
\bibliographystyle{scpma}

\end{document}